\documentclass[conference]{IEEEtran}

\makeatletter

\def\ps@headings{%

\def\@evenhead{\scriptsize\thepage \hfil \leftmark\mbox{}}%

\def\@oddfoot{}%

\def\@evenfoot{}}

\makeatother

\usepackage{algorithmic}

\usepackage{algorithm}
\usepackage{verbatim}
\usepackage{amsmath}
\usepackage{enumerate}
\usepackage{fancybox}

\usepackage{verbatim}
\usepackage{array}
\usepackage{graphicx}
\usepackage{float}
\usepackage[footnotesize]{subfigure}
{\begin{Sbox}\begin{minipage}}%
{\end{minipage}\end{Sbox}\fbox{\TheSbox}}

\begin{document}

\title{Efficient Multicasting in Content-Centric Networks 
Using Datagrams }

\author{J.J. Garcia-Luna-Aceves$^{1,2}$
and Maziar Mirzazad Barijough$^2$ \\
$^1$Palo Alto Research Center, Palo Alto, CA 94304 \\
$^2$Department of Computer Engineering,
 University of California, Santa Cruz, CA 95064\\
 Email: jj@soe.ucsc.edu, maziar@soe.ucsc.edu }

\maketitle

\begin{abstract}

The  Named Data Networking (NDN) and Content-Centric Networking (CCNx) architectures are the leading 
approaches for content-centric networking, and both require using Interests (requests that elicit content) and  maintaining 
per-Interest forwarding state in Pending Interest Tables (PIT) to store per-Interest forwarding state.
To date, PITs have been assumed to be necessary to enable native support for multicasting in the data plane, such 
that multicast forwarding trees (MFT) are established by the forwarding and aggregation of Interests using PITs.
We present a new approach to content-centric networks based on anonymous datagrams that provides native support for multicasting, but does so without  the need to maintain per-Interest forwarding state. 
Simulation experiments  are used to show that the proposed new approach attains the same end-to-end delays for multicasting while requiring orders of magnitude fewer forwarding entries.

\end{abstract}


\section{Introduction}
 
The leading approach to content-centric networking is Interest-based and 
consists of: populating forwarding information bases (FIB) maintained by routers with routes to name prefixes denoting content, sending content requests (called Interests) for specific content objects (CO) over paths implied by the FIBs, and delivering data packets with content objects along the reverse paths traversed by Interests. 

The advantages  that this  approach offers
compared to existing protocols used in the  IP Internet  
include: (a) content providers and caching sites do not know the identity of the consumers requesting content; (b) content can be obtained by name  from 
those sites that are closer to consumers; (c) data packets carrying content cannot traverse loops, because they are sent over the reverse paths traversed by Interests; (d) content-oriented security mechanisms can be implemented as part of the content delivery mechanisms; and (e) supporting multicast traffic is simple to do without additional control signaling.

Named data networking (NDN) \cite{ndn} and CCNx \cite{ccnx} are the two prominent   Interest-based content-centric networking  approaches.
Routers in NDN and CCNx  maintain  per-Interest forwarding state by means of Pending Interest Tables (PIT). The PIT of a router maintains information regarding the incoming interfaces from which Interests for a CO were received and the interfaces where the Interest for the same CO was forwarded. 

Since the introduction of NDN and CCNx, using PITs has been considered a necessity for any Interest-based content-centric networking approach. 
PITs are viewed as essential in order to maintain routes to the origins of Interests while preserving the anonymity of those sources, attain efficient Interest forwarding by aggregating Interests requesting the same content, and support  multicasting without additional support in the control plane. 
Unfortunately, using  PITs comes at a  big price.  PITs grow very large   \cite{dai-12, tsi14, var-13} as the number of Interests from users increases, which is unavoidable 
given that  PITs store per-Interest forwarding state. Just as important,  PITs 
make routers vulnerable to  Interest-flooding attacks \cite{DDos1, vir-13, wahl13a, wahl13b} in which adversaries send malicious Interests that make the size of PITs explode.  Adding to this,  we have shown  \cite{ali-ifip16, icnc16} that  in-network caching obviates the need to use PITs to aggregate Interests as they are forwarded, because the likelihood of Interest aggregation by relaying content routers is minuscule.

We have recently  proposed CCN-GRAM (Gathering of Routes for Anonymous Messengers)  \cite{ccngram15, ifip16} as an 
approach to Interest-based content-centric networking that eliminates the performance limitations resulting from using  PITs in NDN and CCNx. We have shown that CCN-GRAM  \cite{ifip16} substantially outperforms NDN by requiring orders of magnitude smaller forwarding state to deliver content while attaining similar end-to-end delays.
This paper shows that CCN-GRAM supports real-time multicasting of content efficiently using the same Interest-based signaling in the data plane advocated in  NDN and CCNx, but without the need to maintain per-Interest forwarding state. As a result, CCN-GRAM is more efficient than NDN in supporting real-time multicasting of content.


The rest of this paper is organized as follows.
Section \ref{sec-prior} summarizes prior work focusing on the support of multicasting in content-centric networks.
Section \ref{sec-design}  how CCN-GRAM supports the multicasting of content. Section \ref{sec-perf} presents the results of simulation experiments comparing the performance of NDN and CCN-GRAM under multicast traffic.
The results show that CCN-GRAM attains even smaller end-to-end latencies than NDN in retrieving content; however, CCN-GRAM  requires an average number of forwarding entries  per router
that is more than 150 times smaller than the number of PIT entries needed in NDN.

\section{Related Work }
\label{sec-prior}

Prior work on multicasting in content-centric networks is limited.
A number of ICN (information-centric networking) architectures that are not based on Interests adopt a push-based approach to multicasting using mechanisms that are much the same as those  introduced for PIM-SM \cite{pim} for the IP Internet. A good example of this case is COPSS \cite{copss}. Users subscribe to content on a content descriptor (CD), which can be any legal content name, and each CD is associated with a Rendezvous Point (RP). The number of RPs may be as large as the number of ICN nodes. Routers maintain CD-based subscription tables to provide the same functionality as IP multicast, and COPSS supports  sparse-mode multicasting at the content layer. The RP's receive content from one or more publishers and send it over the multicast trees established by routers for the multicast groups.

A major selling point for maintaining per-Interest forwarding state using PITs  in CCNx  and NDN \cite{ccnx, ndn} has been that it enables ``native" support for multicasting in the data plane  with no additional signaling required in the control plane.  
In short, multicast receivers simply send Interests towards the multicast source.
As Interests from receivers and previous-hop routers are aggregated in the PITs  on their way to the multicast source, a multicast  forwarding tree (MFT)  is formed and maintained in the data plane. Multicast Interest are forwarded using the same forwarding information base (FIB) entries used for unicast traffic,
and  multicast data packets are sent using  reverse path forwarding (RPF)  over the paths traversed by aggregated Interests. 
If Interests can be forwarded without incurring loops, then MFTs can be created and multicast data can be disseminated without 
requiring the use of complex multicast routing protocols operating in the control plane (e.g., \cite{pim, mip}.
Using PITs is appealing in this context; however, 
we show below that native  support of multicasting  in the data plane
can be easily done using anonymous datagrams.

\section{Multicasting in CCN-GRAM }
\label{sec-design}

\subsection{Preliminaries}

CCN-GRAM assumes that Interests are retransmitted only by the consumers that originated them, and that  routers use exact Interest matching.
A router that advertises being an origin of a name prefix stores  all the content objects associated with that prefix at a local content store.
Routers know which interfaces are neighbor routers and which are local users within a finite time, and forward Interests on a best-effort basis. For convenience, it is assumed that a request for content from a local user is sent to its local router in the form of an Interest.

The name of content object (CO)  $j$ is denoted by  $n(j)$
and the name prefix that is the best match for  name $n(j)$ is denoted by $n(j)^*$.
Similarly, the name of a multicast group $j$ is denoted by $g(j)$ and the name prefix that is the
best match for  name $g(j)$ is denoted by $g(j)^*$.
The set of neighbors of router $i$ is denoted by $ N^i$.


\subsection{Information Exchanged and Stored}
\label{sec-info}

Like NDN and CCNx, CCN-GRAM uses Interests, data packets, and replies  to Interests. 
Routers differentiate between unicast and multicast Interests. A unicast Interest requests a content object (CO) by name, while a multicast Interest requests content from a multicast group by name.

A multicast Interest $MI[g(j), D^I(i), mc^I(i)]$ sent by router $i$ to router $n$  states: the name of the multicast group $g(j)$ from which content is requested, 
the distance from router $i$  to the source of the multicast group ($D^I(i)$), and a multicast counter ($mc^I(i)$) used for pacing. 

A multicast data packet   $MP[g(j),  sp(j), mc^R(i)]$ states the name of the multicast group $g(j)$ from which content is being sent, 
a security payload ($sp(j)$) used optionally to validate the CO, a multicast counter ($mc^R(i)$), 
plus the content object (CO) corresponding to the value of the multicast counter. 

A  reply sent by router $i$ in response to  a multicast Interest is denoted by
$MR[g(j), CODE, mc^R(i)]$ and contains 
the name of the multicast group for which the Interest was sent, 
a code ($\mathsf{CODE}$) indicating the reason why the reply is sent, and the current value of the multicast counter stored at the router sending the repoy.
Possible reasons for sending a reply include: an Interest loop is detected, or no route is found towards requested content.

Router $i$ uses four tables to forward multicast traffic:
a forwarding information base ($FIB^i$), a multicast anonymous routing table ($MART^i$),   and a group membership table ($GMT^i$).

$FIB^i$ is indexed using known content name prefixes.  
The entry for name prefix $g(j)^*$ states the  distance reported by each next-hop neighbor router for the prefix. The distance stored for  
neighbor $q$ for name prefix $g(j)^*$ in $FIB^i$  is denoted by $D(i, g(j)^*, q)$.
Each entry in $FIB^i$ is stored for a maximum time determined by the lifetime of the corresponding entry in the routing table of the router.


$MART^i$ maintains forwarding state to the receivers of  multicast groups. 
Each entry of the MART specifies a multicast group name, 
the current value of the multicast counter ($mc$) for the group, and a list 
$NH$ of next hops to the group of receivers who have sent multicast  Interests for the group.  
$MART^i [g(j), mc, NH]$ denotes the  entry for group $g(j)$ in $MART^i$.
$GMT^i$ lists the mappings of multicast group names to the lists of local receivers that requested to join the groups. If in-network caching is used as part of multicsting, the entry for group $g(j)$ also states a pointer $p[g(j)]$ to the content 
that has been cached for the group listing each CO by the value of the
multicast counter used to retrieve COs for  $g(j)$.




\subsection{Multicast Content Dissemination}
\label{mint}

Multicast content dissemination is based on the forwarding of Interests along multicast forwarding trees (MFT) to the sources of multicast groups, followed by the forwarding of multicast data packets on the reverse paths traversed by Interests.
Forwarding multicast Interests is based on the information stored in the FIBs maintained by routers.  In contrast to NDN, CCN-GRAM 
maintains forwarding state for Interests on a per-group basis rather than on a per-Interest basis.

Algorithms~\ref{algo-CCN-GRAM-create-Interest}  to \ref{algo-CCN-GRAM-Data}  outline  
the steps  taken by  routers to process and forward multicast Interests, and return multicast data packets   or  replies for the case of real-time multicasting.   

To compare multicasting in CCN-GRAM directly with NDN, we assume  pull-based dissemination of real-time multicast content,
such that a single CO is sent   to multicast receivers in response to an Interest sent  to the source of a multicast group over the MFT.

We assume that each router is initialized properly,  knows the identifiers used to denote local consumers, and knows all its neighbors.
We assume that a routing protocol (e.g., DCR \cite{dcr, gold}) operating in the control plane updates
the entries of  routing tables listing one or multiple next hops towards name prefixes.
Routers populate their FIBs with routes to name prefixes
based on  the data stored in their routing tables. 

The value of the multicast counter for group $g(j)$ in $MART^i$ is denoted by 
$mc^i[g(j)]$, and the set of next-hop routers listed in $MART^i$ for receivers in $g(j)$  is denoted by $NH^i[g(j)]$. The local receivers for group $g(j)$  listed in  $GMT^i$ is denoted by  $GMT^i[g(j)]$.

The forwarding of multicast Interests towards the sources of multicast groups is assumed to rely on the selection of next hops towards name prefixes listed in FIBs that provide the best matches to the multicast group names stated in  the Interests. We assume that routers with local consumers maintain caches of multicast content. 
The first content object (CO) of a multicast group is labeled by the name of the group and a multicast counter equal to one, and an empty entry for a multicast group is initialized with a multicast-counter value equal to zero.
We assume that all  initial requests to join a group state a multicast counter equal to one, and that  forwarding state for a group stored in the MART of  a router is deleted after a timeout if no Interests are received for the group.

For simplicity,  we do not include the steps taken by routers to respond to the failures or additions of interfaces with neighbor routers or local consumers. Furthermore, we assume that Interests and responses to them are transmitted reliably between any two neighboring routers.  In essence, forwarding state related to a failed interface must be deleted and the corresponding replies with negative acknowledgments must be sent to previous next hops to remote receivers or local receivers as needed. Forwarding state associated with new interfaces is instantiated as a result of new Interests being forwarded.

Algorithm \ref{algo-CCN-GRAM-create-Interest}  shows the steps taken by router $i$ to process Interests received from local consumers. 
For convenience, multicast content requests from local consumers are assumed to be Interests stating  the name of a group,   the name of the consumer, and an empty distance to the content assumed to denote infinite. The same format of data packets and replies used among routers is used to denote the responses a router sends to local consumers. Consumers increase  the values of their multicast counters by one  to request the next pieces of multicast content from multicast sources.

Router $i$ adds consumer $c$ as a local receiver in group 
$g(j)$ by adding an entry for $g(j)$ in $GMT^i$ with $c$ as a local receiver for the group, and indicates that it has local receivers  in $MART^i$ by adding itself as a next hop towards receivers of the group.
Router $i$  forwards a single copy of a multicast Interest requesting more content from a multicast source independently of how many local receivers or neighbor routers send multicast Interests to router $i$. This is done by means of the multicast counter ($mc$) maintained by each router and multicast receiver, and the multicast-counter field included in Interests and responses to them.

A content consumer $c$ asks to join a multicast group $g(j)$ as a receiver by sending an Interest $MI[g(j),  $ $D^I(c) = nil, mc^I(c) = 1 ]$. If the value of the multicast counter for the group stored by the router is larger, the router 
responds with the latest multicast data packet corresponding to the current value of the multicast counter maintained by the routers for the multicast group. A router sends a  negative acknowledgment to an Interest from a local consumer
with a  multicast-content value different  than the next expected value to force a retransmission and 
keep all local consumers in the same multicast group using the same current value of the  multicast counter, while  
reducing end-to-end latencies incurred in delivering multicast content to 
consumers far away from group sources.
A consumer requests more content from a multicast group by sending a multicast Interest after incrementing the value of the multicast counter for the group.
 
A router forwards Interest $MI[g(j), D^I(i), mc^I(i)]$ 
towards the source of multicast group $g(j)$ based on the information in its FIB.

\begin{algorithm}[h]
\caption{Processing Interest  from user $c$ at router $i$}
\label{algo-CCN-GRAM-create-Interest}
{\fontsize{7}{7}\selectfont
\begin{algorithmic}
\STATE{{\bf function}  Interest\_Source}
\STATE {\textbf{INPUT:}  $GMT^i$,   $FIB^i$, $MART^i$,  $MI[g(j),  D^I(c) = nil, mc^I(c)]$}

\IF{$g(j)^* \in FIB^i$    ~(\% Route to $g(j)$ exists)}


	\IF{$MART^i$ entry for $g(j)$ does not exist}
		\STATE{
		$mc^i[g(j)] = 0$;  $NH^i[g(j)] = \emptyset$; }
		\STATE{create entry $MART^i [g(j), mc, NH ] $; $GMT^i[g(j)] = \emptyset$;    }
	\ENDIF
	\STATE{$GMT^i[g(j)] = GMT^i[g(j)]   \cup c$;  
	$NH^i[g(j)] = NH^i[g(j)] \cup i$;}

	\IF{$mc^I(c) \not=  mc^i[g(j)] +1$}
			\IF{$p[g(j)] \not= nil$ }
				\STATE{
				retrieve CO for $mc^i[g(j)]$; 
				$mc^R(i) = mc^i[g(j)]$; \\
				send $MP[g(j),  sp(j), mc^R(i) ]$ to $c$}
			\ELSE
				\STATE{$mc^R(i) = mc^i[g(j)]$; 
				send $MR[g(j), \mathsf{Interest~error}, mc^R(i) ]$ to $c$}
			\ENDIF
	\ELSE 
			\IF{ $i$ is the source for $g(j)$ }
				\STATE{
				$mc^R(i) = mc^i[g(j)]$;}
				\STATE{send $MP[g(j),  sp(j), mc^R(i) ]$ \\ to receivers in $GMT^i[g(j)]$ and next hops in $NH^i[g(j)]$}
			\ELSE	
				\STATE{ $mc^I(i) = mc^i[g(j)]$; 
				}
				\STATE{
				$D^I(i) = Min \{ D(i, g(j)^*, u) $ for $ u \in S^i_{g(j)^*} \}$;}
				\FOR{{\bf each} $v \in N^i$ {\bf by rank in} $FIB^i$} 
					\IF{$D(i, g(j)^*, v) = D^I(i)$}
					\STATE{
					send $MI[g(j), D^I(i), mc^I(i) ]$ to  $v$;  {\bf return}	}
					\ENDIF
				\ENDFOR	
			\ENDIF

	\ENDIF
\ELSE
	\STATE{$mc^R(i) = mc^i[g(j)]$;  send $MR[g(j),  \mathsf{no~ route},  mc^R(i) ]$ to $c$ } 

\ENDIF

\end{algorithmic}
}
\end{algorithm}

 \vspace{-0.05in}
Algorithm~\ref{algo-CCN-GRAM-Interest} shows the steps taken by router $i$ to process an Interest received from a neighbor router $p$. 
Router $i$  follows similar steps to those in Algorithm 1 to respond to an Interest with a multicast  data packet to the neighbor router  if the content is local and the multicast counter in the Interest is smaller than the current value of the multicast counter at the router. 
If the Interest requests the next CO from the group and the group source is local, the multicast data packet is sent to all next hops along  the MFT. Alternatively, if  
the multicast source is remote, the router forwards the Interest ensuring that no forwarding loops occur.
Let $S^i_{g(j)^*}$ denote the set of next-hop neighbors of router $i$ for prefix $g(j)^*$. The  following  rule  is used to 
ensure that multicast Interests cannot traverse routing loops, even if the 
routing data stored in FIBs regarding name prefixes is inconsistent and leads to routing-table loops.

 \vspace{0.05in}
\noindent
{\bf  Loop-Free Forwarding Rule (LFR):}  \\
Router $i$  
accepts $MI[n(j),  D^I(k), mc^I(k) ]$ from router $k$ if:
\begin{equation}
\label{lfr}
\exists ~v \in  S^i_{g(j)^*} (~ D^I(k)  > D(i, g(j)^*, v) ~) 
\end{equation}

Router $i$ tries to forward the Interest to a next hop $s$ for the best prefix match for $n(j)$ that satisfies LFR. 
The highest-ranked router 
satisfying LFR is selected as the successor for the Interest and router $i$.
If no neighbor is found that satisfies LFR, a reply is sent stating that a loop was found.

LFR is based on the same approach   proposed previously to
eliminate Interest looping in NDN and CCNx  \cite{ifip2015, ancs2015}. 
It ensures loop-free forwarding of Interests by ensuring that routers forward Interests only to next hops that are 
closer to the intended name prefix.

\begin{algorithm}[h]
\caption{Processing multicast Interest  from router $p$ at router $i$}
\label{algo-CCN-GRAM-Interest}
 {\fontsize{7}{7}\selectfont
\begin{algorithmic}
\STATE{{\bf function} Interest\_Forwarding}
\STATE {\textbf{INPUT:}  $GMT^i$, $FIB^i$, $MART^i$,  $MI[g(j), D^I(p), mc^I(p)]$;}

\IF{$g(j)^* \in FIB^i$    ~(\% Route to $g(j)$ exists)}


	\IF{$MART^i$ entry for $g(j)$ does not exist}
		\STATE{
		$mc^i[g(j)] = 0$;  $NH^i[g(j)] = \emptyset$; \\
		create entry $MART^i [g(j), mc, NH  ] $; $GMT^i[g(j)] = \emptyset$ }
	\ENDIF

		\STATE{ $NH^i[g(j)] = NH^i[g(j)] \cup p$; }  
		\IF{$mc^I(p) \not= mc^i[g(j)] +1$}
			\STATE{$mc^R(i) = mc^i[g(j)]$; }
			\IF{$p[g(j)] \not= nil$ }
				\STATE{
				retrieve CO for $mc^i[g(j)]$; 
				send $MP[g(j),  sp(j), mc^R(i) ]$ to $p$}
			\ELSE
				\STATE{send $MR[g(j), \mathsf{Interest~error}, mc^R(i) ]$ to $p$}
			\ENDIF
		\ELSE
			\STATE{
			$mc^i[g(j)] = mc^i[g(j)] + 1$; }
			\IF{ $i$ is the source for $g(j)$ }
				\STATE{ retrieve CO for $mc^i[g(j)]$; $mc^R(i) = mc^i[g(j)]$; \\
				send $MP[g(j), sp(j), mc^R(i) ]$  to \\
				receivers in $GMT^i[g(j)]$ 
				and next hops in $NH^i[g(j)]$}
			\ELSE	
				\STATE{ 
				$D^I(i) = Min \{ D(i, g(j)^*, u)$ for $ u \in S^i_{g(j)^*} \}$;}
				\FOR{{\bf each} $v \in N^i$ {\bf by rank in} $FIB^i$} 
					\IF {$ D^I(p)  > D^I(i) $   ~(\% LFR is satisfied) }
						\STATE{ $mc^I(i) = mc^i[g(j)]$;   send $MI[g(j), D^I(i), mc^I(i) ]$ to  $v$;  }
						\STATE{{\bf return}	}
					\ENDIF
				\ENDFOR	
				\STATE{ $mc^R(i) = mc^i[g(j)]$; send  $MR[g(j),  \mathsf{loop}, mc^R(i) ]$ to $p$}
			\ENDIF


		\ENDIF

\ELSE
	\STATE{$mc^R(i) = mc^i[g(j)]$; send $MR[g(j),  \mathsf{no~ route},  mc^R(i) ]$ to $p$ } 

\ENDIF

\end{algorithmic}
}
\end{algorithm}

Algorithm~\ref{algo-CCN-GRAM-Data} outlines the processing of  multicast data packets.  
If  local consumers requested the content in the data packet, it is sent to those consumers based on the information stored in $GMT^i$. If the router has neighbor routers that are next hops towards remote receivers of the multicast group, router $i$ forwards the data packet to all neighbors listed for $g(j)$ in $NH^i[g(j)]$ other than router $i$ itself if there are local receivers.
Routers take similar steps in the forwarding of replies to multicast Interests when retransmissions are done by consumers, i.e., routers simply forward replies back to the consumers along the MFT created by the forwarding of multicast Interests.

\begin{algorithm}[h]
\caption{Processing multicast data packet from router $s$ at router $i$}
\label{algo-CCN-GRAM-Data}
{\fontsize{7}{7}\selectfont
\begin{algorithmic}
\STATE{{\bf function} Multicast Data Packet}
\STATE{\textbf{INPUT:}  $GMT^i$,  $MART^i$, 
$MP[g(j),  sp(j), mc^R(s) ]$; }
\STATE{{\bf [o]} verify $ sp(j)$;}
\STATE{{\bf [o]} {\bf if} verification with $ sp(j)$ fails {\bf then} discard $MP[g(j),  sp(j), mc^R(s) ]$;}

\IF{$NH^i[g(j)] \not= \emptyset$}
	\STATE{
	$ mc^R(i)= mc^R(s)$; {\bf if} $mc^i[g(j)] < mc^R(s)$ {\bf then} $mc^i[g(j)] = mc^R(s)$;  }
	\IF{ $GMT^i[g(j)] \not= \emptyset$~~(\% router $i$ has local receivers in group $g(j)$) }
		\FOR{{\bf each} $c \in GMT^i[g(j) $}
			\STATE{send $MP[g(j), sp(j), mc^R(i) ]$ to  $c$}
		\ENDFOR
	\ENDIF
	\IF{$NH^i[g(j)] - \{ i \} \not= \emptyset$}
		\FOR{{\bf each} $h \in NH^i[g(j)]  - \{ i \} $}
			\STATE{send $MP[g(j), sp(j), mc^R(i) ]$ to  $h$}
		\ENDFOR
	\ENDIF
	\STATE{{\bf[o]} store CO in local storage at $p[g(j)]$ indexed with $mc^R(i)$}
\ELSE
\STATE{drop $MP[g(j),  sp(j), mc^R(s) ]$}
\ENDIF

\end{algorithmic}}       
\end{algorithm}

\vspace{-0.1in}
\section{Example of Multicast Dissemination in CCN-GRAM}
\label{sec-mcast}

\begin{figure}[h]
\begin{centering}
    \mbox{
    \subfigure{\scalebox{.18}{\includegraphics{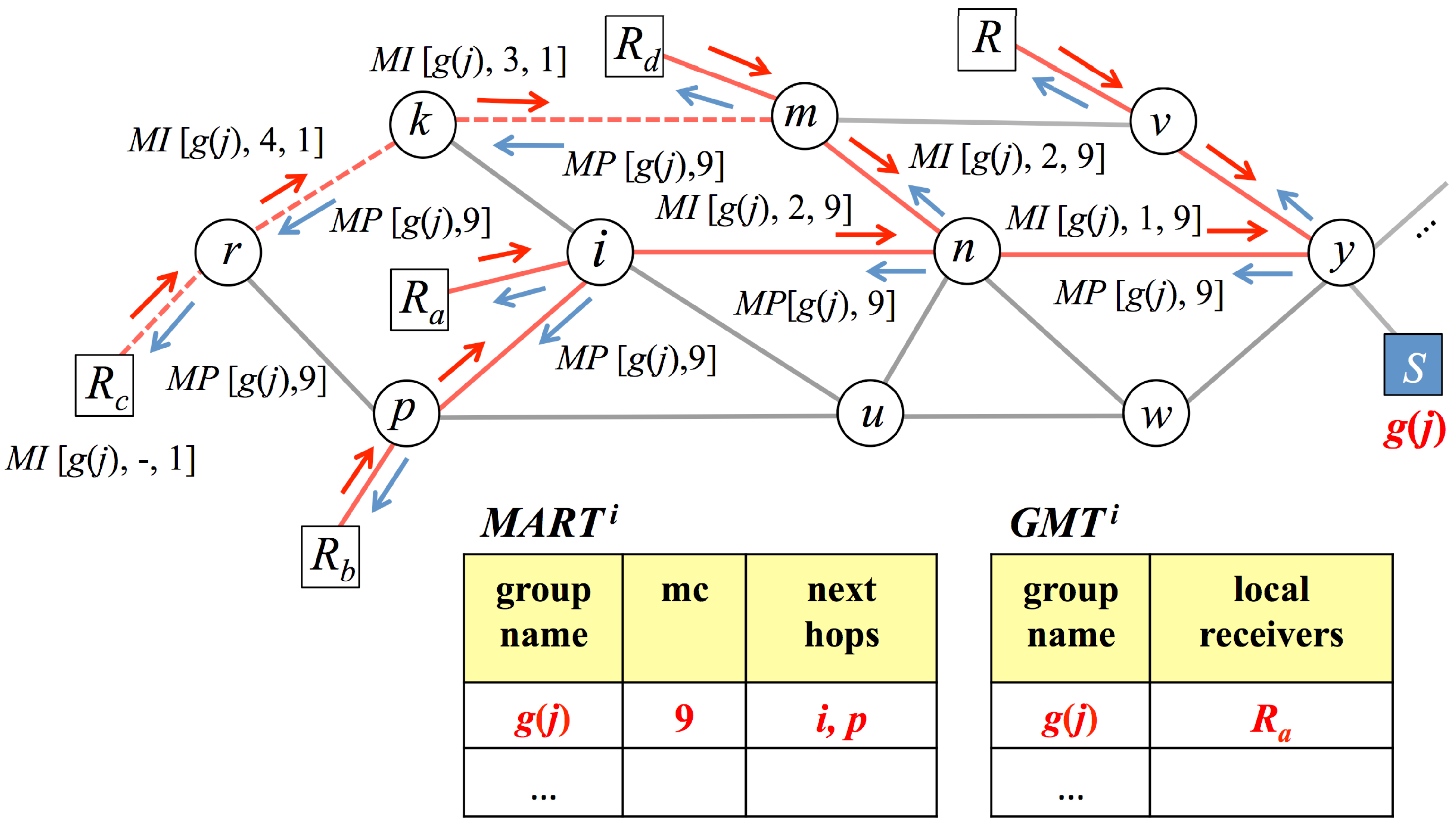}}}
      }
 \vspace{-0.08in}
   \caption{Native multicast support in CCN-GRAM
   }
   \label{gram-example}
\end{centering} 
\end{figure}

Figure \ref{gram-example} illustrates the forwarding of multicast Interests and multicast data packets in CCN-GRAM. As the figure shows, router $i$ maintains a forwarding table ($MART^i$) specifying the next hops to multicast receivers for each multicast-group name, and a table ($GMT^i$) listing the local receivers for each multicast-group name. As the figure shows, the entry for group $g(j)$ in $MART^i$ lists router $i$ as a next hop, which indicates the presence of local receivers; the one local receiver ($R_a$) for group $g(j)$ is listed in $GMT^i$.

\vspace{-0.2in}

The entries in the MARTs and GMTs maintained by routers define the forwarding multicast trees (FMT) of all multicast groups created in the network, and are established by the forwarding of Interests, just as in NDN or CCNx. 
However, the use of multicast counters eliminates the need to maintain per-Interest forwarding state.
In the figure, dashed lines represent links along the path from consumer $R_c$ joining the multicast group after the source has disseminated CO with $mc = 9$ to the rest of the MFT. The late joiner is brought up to the current state of the multicast group  by the
multicast counter carried in each data packet.

\section{Performance Comparison}
\label{sec-perf}

We compare the average table sizes and end-to-end delays for multicast traffic in CCN-GRAM and NDN  by running experiments based on implementations of CCN-GRAM and NDN in the ndnSIM simulation tool \cite{ndnsim}. The implementation of CCN-GRAM is based on the algorithms  presented in this paper, and NDN implementation from ndnSIM is used without modification. The simulation scenario includes 200 nodes distributed uniformly in a $100m \times 100m$ area. Nodes with a distance  of 15 meters or less from each other are connected with a point-to-point link of 15ms delay. Data rates are set to 1Gbps to eliminate or reduce the impact  of an inefficient implementation of either NDN and CCN-GRAM on the results. Using on-path caching strategy, each router in these experiments can cache up to 1000 content objects. In each simulation scenario, there are multiple multicast groups, and  each group contains multiple consumers and one producer. Consumer and producer nodes for each group is selected at random from the 200 routing nodes in the network.

We compared forwarding table  sizes in NDN and CCN-GRAM by four different varying parameters: Multicast groups count, multicast group size, Interest request rate, and link delay. We also compared average end-to-end delay in NDN and CCN-GRAM for different request rates. For this purpose we consider minimum download rate of 1.5 Mbps for audio/radio streaming, 5 Mbps for HD video streaming, and 25Mbps for Ultra HD video streaming. Considering  the standard packet size of 4KB  advocated in NDN, we compared different scenarios with constant rate of 50 to 800 interests per second from  each consumer application. 

\subsection{Size of Forwarding Tables}

\begin{figure}[h]
\begin{centering}
    \mbox{
    \subfigure{\scalebox{.68}{\includegraphics{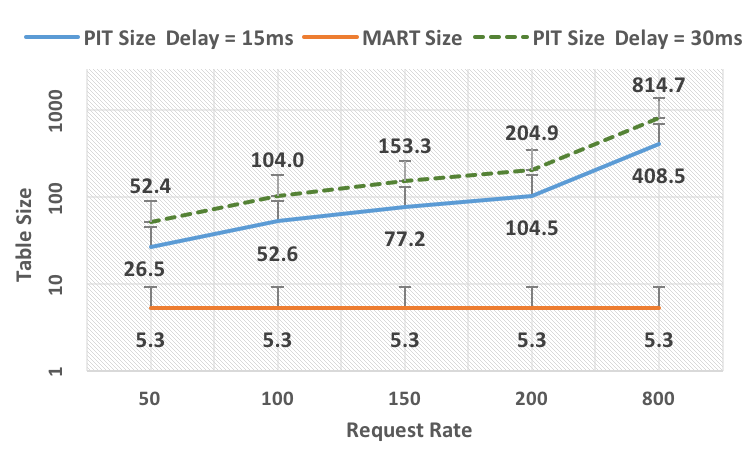}}}
    }
\vspace{-0.21in}
   \caption{Average size of forwarding tables for varying request rates }
   \label{rate}
\end{centering} 
\end{figure}

Figure \ref{rate} shows the results of a simulation experiment that includes  20 multicast groups, each with 20 consumers and one producer. 

\vspace{-0.22in}
\vspace{-0.05in}

The above figure shows the average size of a forwarding table in logarithmic scale as a function of Interest (request) rates.
Given that CCN-GRAM adds a single entry per multicast group, the number of entries in  a MART is independent of request rate.  By contrast, the size of PITs in NDN is a function of the rate at which Interests arrive at routers. 
The maximum  MART size for this scenario is 20, and the average size of a MART table is 5.26 independently of the  request rates. As the figure shows, the  number of PIT entries is highly affected by the Interest rates from consumers. For the case of a 15ms link delay, increasing the Interest rate to 800 results in average PIT size of 408 and tables as large as 1300 entries for routers.

\vspace{-0.22in}
\begin{figure}[h]
\begin{centering}
    \mbox{
    \subfigure{\scalebox{.70}{\includegraphics{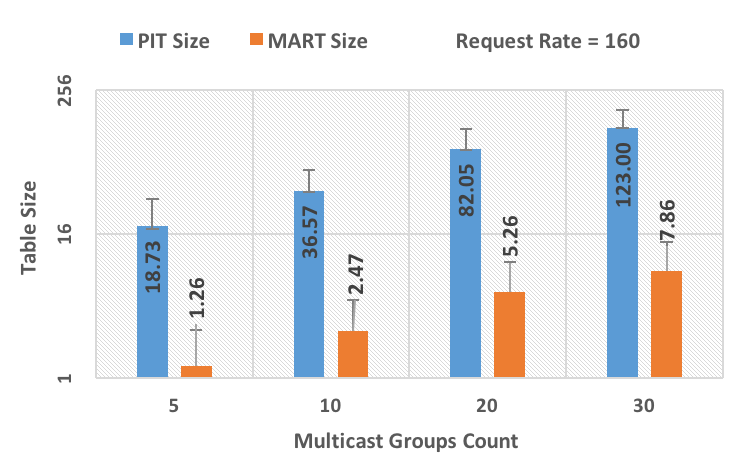}}}
    }
\vspace{-0.22in}
   \caption{Average size of forwarding tables vs. number of multicast groups}
   \label{mcgCount}
\end{centering} 
\end{figure}  

Figure \ref{mcgCount} shows the average MART size versus the average PIT size for varying number of multicast groups from 5 to 30 groups, with each group having 20 consumers with Interest (request)  rate of 160 Interests per second, which is enough to support  HD video streaming with each data packet being 4KB. In CCN-GRAM,  the number of entries of MART tables cannot exceed the total  number of multicast groups. On the other hand, as the figure shows,  the number of entries in the PIT of a router  is directly related to the number of interests 
received by the router, which in turn depends on the number multicast groups and the request rate per group.
Accordingly, the average PIT size can grow dramatically.

\vspace{-0.2in}
\begin{figure}[h]
\begin{centering}
    \mbox{
    \subfigure{\scalebox{.70}{\includegraphics{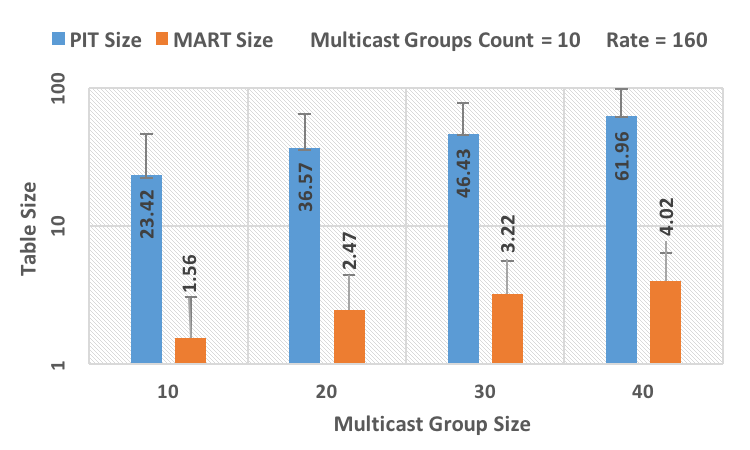}}}
    }
\vspace{-0.25in}
   \caption{Average size of forwarding tables for varying size of multicast groups}
   \label{mcgSize}
\end{centering} 
\end{figure}

Figure \ref{mcgSize} shows the  average size of PITs and MARTs  for varying multicast group sizes from 10 to 40 consumers per group.
As the size of a multicast group increases, more routers become involved in forwarding multicast Interests and multicast data packets in both NDN and CCN-GRAM. This results in larger  average  sizes of both PITs and MARTs. However, the grow rate for NDN is higher because of entries are added to PITs on a per Interest basis. 

\subsection{Average Delays}

\begin{figure}[h]
\begin{centering}
    \mbox{
    \subfigure{\scalebox{.70}{\includegraphics{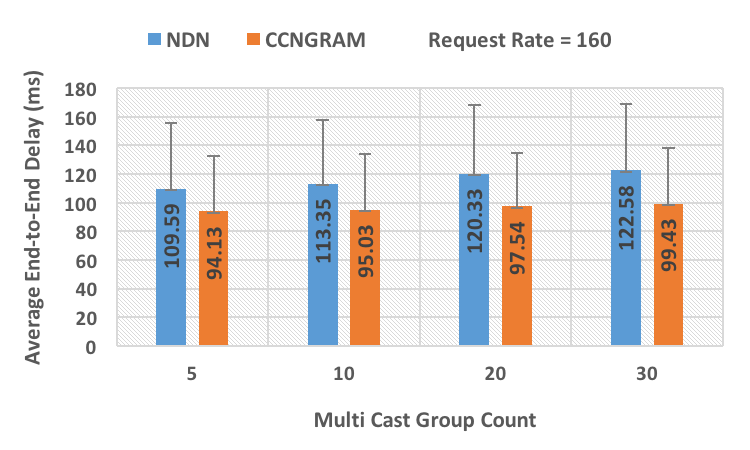}}}
    }
\vspace{-0.26in}
   \caption{Average end-to-end delay for varying number of multicast groups}
   \label{delay}
\end{centering} 
\end{figure}

As Figure \ref{delay} shows, the average delay for CCN-GRAM is shorter than the delays incurred in NDN. 
According to Algorithm 3,  the first multicast Interest received by the producer, results in multicast of data toward current members of multicast group in CCN-GRAM, even if the Interest from a member or previous hop relay in the MFT has not been received yet.  On the other hand, in NDN,  if one consumer node is far from the producer compared to other consumer nodes such that its interests is not aggregated with the same interests from other consumers, request of that node for a multicast data will be processed separate from other group members, which results in lower throughput and higher delays. The operation of NDN could  be modified to
mimic the way in which CCN-GRAM forwards multicast data over MFTs, in which case end-to-end latencies would be similar.

\vspace{-0.2in}

The results shown in Figures \ref{mcgCount}, \ref{mcgSize}, and \ref{delay} are for link delays of 15ms, which results in end-to-end delays of around 110 ms. However if  higher link delays occur, the average number of PIT entries can grow to much larger values, 
because the number of pending interests in each router increases. As shown in Figure \ref{rate},  link delays  of 30ms  results in 240 end-to-end delays and much larger PITs. For the case of 800 Interests per second, the average PIT size in NDN  increases to 814 entries 
some relaying routers storing as many as as 2700 PIT entries. On the other hand, the average size of MARTs is not affected by varying link delays and remains constant, which makes CCN-GRAM more scalable and reliable, and more attractive for real-time streaming applications requiring 
high throuhput.

\section{Conclusions and Future Work}

We presented CCN-GRAM, the first approach for multicasting in content-centric networks that eliminates the need to maintain per-Interest forwarding state and still
operates based  on the forwarding of Interests in the data plane, without the need for a multicast routing protocol in the control plane.  


Simulation experiments were used to show that the 
storage requirements for  CCN-GRAM are orders of magnitude smaller than for NDN, and that end-to-end delays  in  CCN-GRAM are similar if not smaller than in NDN, especially when high data rates for multicast streaming is needed.

Additional work is needed to define efficient mechanisms to react to resource failures with minimum disruption to MFTs, reliable multicasting  support, and new flow control approaches  that allow more than one CO to be delivered per multicast Interest.
In addition, CCN-GRAM could be applied when sources disseminate content without the need for receivers to submit Interests after joining a group \cite{ifip16}.



 {\fontsize{7}{7}\selectfont

  }

\end{document}